\documentclass[prl,aps,floatfix]{revtex4}
\usepackage{epsfig} \usepackage{graphics} \usepackage{bm}
\usepackage{amssymb}
\usepackage{graphicx}
\begin{document}

\preprint{Lebed-PRB-LN}

\title{Violation of the Einstein's Equivalence Principle for a
Composite Quantum Body}

\author{A.G. Lebed}

\affiliation{Department of Physics, University of Arizona, 1118 E.
4-th Street, Tucson, AZ 85721, USA}

\begin{abstract}
Recently, we have started to investigate behavior of a composite
quantum body in an external gravitational field in the framework
of General Relativity [see, for a review, A. G. Lebed, Mod. Phys.
Lett. A, {\bf 35}, 2030010 (2020)]. As the simplest example, we
have considered a hydrogen atom in a weak gravitational field. Our
results are the following. The Einstein's Equivalence Principle
survives for the most of macroscopic ensembles of the atoms,
containing the stationary quantum states. On the other hand, we
have demonstrated that this principle is sometimes broken. In
particular, it is broken for the so-called Gravitational demons,
which are the coherent macroscopic ensembles of two or more
stationary quantum states in the hydrogen atoms. In the above
cited paper we have considered the Gedanken experiment, where the
gravitational field is suddenly switched on in a free from
gravitation space. In the current paper we consider the much more
realistic from experimental point of view Gedanken experiment and
come to the same conclusion about violations of the Einstein's
Equivalence Principle for the Gravitational demons.
\end{abstract}

\maketitle

The Einstein's Equivalence Principle of equality between inertial
and gravitational masses is a cornerstone of the theory of General
Relativity [1,2]. During the recent space mission "MICROSCOPE",
validity of the principle for ordinary condensed matter was
established with a great accuracy, $|m_i - m_g|/m_i \leq 10^{-16}
- 10^{-17}$ [3,4], whereas two other space missions, "GG" and
"STEP", are currently in their preparation stages. Nevertheless, a
validity of the Einstein's Equivalence Principle in quantum case
is not obvious and needs its theoretical and hopefully
experimental investigations. Since quantum theory of gravity has
not been developed yet, in our previous papers (see Refs. [5-8])
we have theoretically considered the so-called semi-classical
variant of General Relativity, where gravitational field is not
quantized but the matter is quantized. In particular, in the
framework of the semi-classical approach, we have shown that
active and passive gravitational masses for the simplest quantum
macroscopic ensembles - ensembles of hydrogen atoms in stationary
states - preserve the Einstein's Equivalence Principle. On the
other hand, we have also demonstrated that there exist such
unusual quantum ensembles, called the Gravitational demons, which
result in violations of the Einstein's Equivalence Principle for
both active [5,6,8] and passive [6-8] gravitational masses. Note
that the Gravitational demons are defined as the coherent
macroscopic ensembles of the quantum superpositions of two or
several stationary states. As shown by us in the above cited
papers, the expectation values of active and passive gravitational
masses in such macroscopic ensembles are not anymore related to
the expectation value of energy by the famous Einstein's formula,
$m_g \neq E/c^2$. In particular, in Refs. [6-8] we have considered
the Gedanken experiment, where the Gravitational demons are
created by laser technique in ultra-cold matter [9] in the absence
of a gravitational field and then we have switched on the field at
$t=0$. In this paper, we consider more realistic from experimental
point of view Gedanken experiment, where we create the
Gravitational demon in the presence of gravitational field and
then at $t=0$ we change the field a little bit. Our conclusions
regarding passive gravitational mass are similar to the above
discussed ones.

Let us consider for further calculations the so-called weak
gravitational field limit [1,2], which describes the Earth's
gravitational field with great accuracy in the framework of the
General Relativity spacetime,
\begin{equation}
(ds)^2(R) = -[1 + 2 \phi(R)/c^2 ](cdt)^2 + [1 - 2 \phi(R)/c^2 ]
[(dx)^2 +(dy)^2+(dz)^2 ], \ \ \ \ \phi (R)= - GM/R,
\end{equation}
where $\phi(R)$ is the Newtonian potential, $c$ is the light
velocity, $G$ is the Newtonian gravitational constant, $R$ is a
distance from a center of the Earth, and $M$ is the mass of Earth.
Note that, in accordance with the local Lorentz invariance, we can
define the following proper local space-time coordinates,
\begin{equation}
\tilde x(R) =[1-\phi(R)/c^2]  x, \ \ \ \tilde y(R)= [1-\phi(R)/c^2
] y, \ \ \ \tilde z(R)=[1-\phi(R)/c^2 ] z , \ \ \ \tilde t(R)=
[1+\phi(R)/c^2] t,
\end{equation}
where the spacetime interval has the Minkowski's form:
\begin{equation}
[d \tilde s(R)]^2 = -[c d \tilde t(R)]^2 + [d \tilde x(R)]^2 + [d
\tilde y(R)]^2 + [d \tilde z(R)]^2.
\end{equation}
[Note that the weak field approximation (1),(2) allows us below to
perform all calculations with accuracy up to the first order of
the small parameter $|\phi|/c^2 \ll 1$.] The Schr\"{o}dinger
equation for electrons in a hydrogen atom in the proper local
coordinates (2) can be expressed in the approximation, where we
neglect the so-called tidal terms, as
\begin{equation}
i \hbar [\partial \Psi({\bf \tilde r},\tilde t)]/ \partial \tilde
t = \hat H_0 (\hat {  \bf \tilde p},{\bf \tilde r}) \Psi({\bf
\tilde r},\tilde t)  ,
\end{equation}
where $\hat H_0 (\hat {  \bf \tilde p},{\bf \tilde r})$ is the
electron Hamiltonian in the hydrogen atom. We stress that Eq.(4)
can be written if position of the proton is fixed by some
non-gravitational forces and if we omit all the so-called tidal
terms, which are extreemely small near the Earth.

Below, we perform the following Gedanken experiment. At $t=0$ we
create a quantum macroscopic ensemble of the coherent
superpositions of the two stationary states in the atom -- $1S$
(with the corresponding wave function $\Psi_1[{\bf \tilde r(R)}]$)
and $2S$ (with the corresponding wave function  $\Psi_2[{\bf
\tilde r(R)}]$):

\begin{equation}
\tilde \Psi^1_{12}(\tilde r, \tilde t) = \exp [-i m_e c^2 \tilde
t(R) /\hbar ]
 \frac{1}{\sqrt{2}} \biggl\{ \Psi_1[\tilde r(R)] \exp [-i E_1 \tilde t(R)/ \hbar]
+ \Psi_2[\tilde r(R)] \exp(i \tilde \alpha) \exp [-i E_2 \tilde
t(R)/ \hbar] \biggl\} ,
\end{equation}
where a phase difference between the stationary wave functions is
the same for each microscopic state in the macroscopic ensemble,
\begin{equation}
\tilde \alpha = const,
\end{equation}
and the wave functions of the stationary states are known to be
real and are normalized in the local proper coordinates:
\begin{equation}
\int^{+\infty}_{-\infty} [\Psi_1(\tilde r)]^2 d^3 \tilde r =
\int^{+\infty}_{-\infty} [\Psi_2(\tilde r)]^2 d^3\tilde r =1, \ \
\ \ \int^{+\infty}_{-\infty}  \Psi_1(\tilde r) \Psi_2(\tilde r)
d^3 \tilde r=0,
\end{equation}
where $m_e$ is the electron bare mass. For convenience, here we
rewrite wave function (5) in the initial spacetime coordinates,
\begin{eqnarray}
&&\Psi^1_{12}(r,t) = \exp \{ -i m_e c^2 t [1+\phi (R_1) /c^2] /
\hbar \} [1- \phi(R_1) /c^2]^{3/2}
\nonumber\\
&&\times \biggl( \Psi_1[r(1-\phi(R_1)/c^2)]  \exp \{-i E_1
t[1+\phi(R_1)/c^2] / \hbar \}  \
\nonumber\\
&&+ \Psi_2[r(1-\phi(R_1)/c^2)]  \exp(i \tilde \alpha) \exp \{-i
E_2 t[1+\phi(R_1)/c^2] / \hbar\} \biggl) / \sqrt{2} ,
\end{eqnarray}
where $\phi_1 = \phi(R_1)$ is value of the gravitational potential
(1) at initial moment of time, $t=\tilde t =0$.

Let us continue our Gedanken experiment. We suddenly change the
gravitational potential into the value $\phi_2=\phi(R_2) =
\phi(R_1) + \delta \phi$, where $|\delta \phi| \ll |\phi(R_1)|$,
and ask how the expectation value of energy of our Gravitational
demon (5),(8) is changed, provided that time of the change of the
gravitational potential is less than the characteristic time of
the electron quasi-classical motion corresponding to the wave
function (8). Then, at $t > 0$, we have the following wave
function:
\begin{eqnarray}
&&\Psi^2_{12}(r,t) =\exp \{ -i m_e c^2 t [1+\phi (R_2) /c^2] /
\hbar \} [1- \phi(R_2) /c^2]^{3/2}
\nonumber\\
&&\times \biggl(A \ \Psi_1[r(1-\phi(R_2)/c^2)]  \exp \{-i E_1
t[1+\phi(R_2)/c^2] / \hbar \}  \
\nonumber\\
&&+ B \ \Psi_2[r(1-\phi(R_2)/c^2)]  \exp \{-i E_2
t[1+\phi(R_2)/c^2] / \hbar\} \biggl)  ,
\end{eqnarray}
where A and B are some complex numbers. Since we have a sudden
perturbation of the spacetime in our Gedanken experiment, a law of
the conservation of the number of particles can be expressed as
\begin{equation}
|\Psi^1_{12}(r,t=0)|^2  = |\Psi^2_{12}(r,t=0)|^2 .
\end{equation}
[Note that probabilities to occupy higher energy levels are
proportional to $(\delta \phi)^2/c^4$ and, thus, are not
considered in this paper.] It is important that we are not
interested in common phase of the complex numbers $A$ and $B$,
therefore, we can rewrite Eq.(10) in the following form,
\begin{equation}
\Psi^1_{12}(r,t=0) = \Psi^2_{12}(r,t=0).
\end{equation}
Taking into account Eqs.(8) and (9), we obtain
\begin{eqnarray}
&&\frac{1}{\sqrt{2}} \biggl( 1-\frac{\phi_1}{c^2}\biggl)^{3/2}
\biggl\{\Psi_1[r(1-\phi_1/c^2)] \ + \exp(i \tilde \alpha)
\Psi_2[r(1-\phi_1/c^2)]\biggl\}
\nonumber\\
&&=\biggl( 1-\frac{\phi_2}{c^2}\biggl)^{3/2} \biggl\{A \
\Psi_1[r(1-\phi_2/c^2)] \ + B \ \Psi_2[r(1-\phi_2/c^2)]\biggl\} .
\end{eqnarray}

Let us solve Eq.(12) using orthogonality conditions between the
wave functions $\biggl(
1-\frac{\phi_i}{c^2}\biggl)^{3/2}\Psi_1[r(1-\phi_i/c^2)]$ and
$\biggl( 1-\frac{\phi_i}{c^2}\biggl)^{3/2}
\Psi_2[r(1-\phi_i/c^2)]$, where $i=1,2$. As a result we find:
\begin{equation}
\frac{1}{\sqrt{2}} = A \  I_{11}+B \ I_{12}, \ \ \
\frac{\exp(i\tilde \alpha)}{\sqrt{2}} = A \ I_{21} + B \ I_{22} \
,
\end{equation}
where
\begin{equation}
I_{ij}= \biggl(1- \frac{\phi_1}{c^2} \biggl)^{3/2} \biggl(1-
\frac{\phi_2}{c^2} \biggl)^{3/2} \int^{+\infty}_{-\infty}
\Psi_i\biggl[r\biggl(1-\frac{\phi_1}{c^2} \biggl)\biggl] \
\Psi_j\biggl[r\biggl(1- \frac{\phi_2}{c^2} \biggl)\biggl] d^3r.
\end{equation}
Here, we calculate the matrix elements $I_{ij}$ by means of the
method developed in Refs. [6-8] with accuracy up to the first
orders of $|\delta \phi|/c^2 \ll 1$ and $|\phi_i|/c^2 \ll 1$.
After the lengthy but straightforward calculations, we obtain
\begin{equation}
I_{11} = I_{22}  = 1.
\end{equation}
and
\begin{equation}
I_{12} = -  I_{21}  = \frac{\delta \phi}{c^2} \alpha =
\frac{\delta \phi}{c^2} \int^{+\infty}_{-\infty} \Psi'_1(r) \ r \
\Psi_2(r) d^3r =  \frac{\delta \phi}{c^2} \frac{V_{12}}{E_2-E_1} ,
\  \  \ \Psi'_1 (r) = \frac{d \Psi_1(r)}{dr},
\end{equation}
where $E_1$ and $E_2$ are energies of the ground state $1S$ and
the first exited state $2S$, correspondingly;    $V_{12}$ is the
matrix element of the so-called quantum virial operator in the
hydrogen atom [6-8,10]:
\begin{equation}
V_{12}=\int^{+\infty}_{-\infty} \Psi_1(r) \hat V(r) \Psi_2(r)
d^3r, \ \ \ \hat V(r) = \frac{{\bf p}^2}{m_e}-\frac{e^2}{r'},
\end{equation}
where ${\bf p}$ is electron momentum operator, $r'$ is a distance
between electron and the fixed proton; $e$ is the electron charge.

First of all, we solve the system of linear equations (13) to find
the complex numbers A and B:
\begin{equation}
A= \frac{1}{\sqrt{2}} \biggl[ 1 - \alpha \biggl(\frac{\delta
\phi}{c^2} \biggl) \exp(i \tilde \alpha) \bigg], \ \ \ \ B=
\frac{1}{\sqrt{2}} \biggl[ \exp(i \tilde \alpha) + \alpha \biggl(
\frac{\delta \phi}{c^2}\biggl) \biggl] .
\end{equation}
Then, we make sure that they preserve the normalization conditions
with the accuracy of calculations accepted in this paper,
\begin{eqnarray}
|A|^2 + |B|^2 = \frac{1}{2} \biggl\{ \biggl[ 1 - \alpha \biggl(
\frac{\delta \phi}{c^2} \biggl) \exp(i \tilde \alpha) \biggl]
\biggl[ 1 - \alpha \biggl( \frac{\delta \phi}{c^2} \biggl) \exp(-
i \tilde \alpha) \biggl]
\nonumber\\
+\biggl[ \exp(i \tilde \alpha) + \alpha \biggl( \frac{\delta
\phi}{c^2}\biggl) \biggl]  \biggl[ \exp(- i \tilde \alpha) +
\alpha \biggl( \frac{\delta \phi}{c^2}\biggl) \biggl] \biggl\}
\approx 1 .
\end{eqnarray}
And finally, we calculate the expectation energy of the
Gravitational demon (9):
\begin{eqnarray}
\biggl< E \biggl> = \biggl[ E_1 + m_e \delta \phi +
\biggl(\frac{E_1}{c^2}  \biggl) \delta \phi \biggl] |A|^2 +
\biggl[ E_2 + m_e \delta \phi + \biggl(\frac{E_2}{c^2}  \biggl)
\delta \phi \biggl] |B|^2
\nonumber\\
 = \frac{1}{2} \biggl\{ \biggl[ E_1 + m_e \delta \phi + \biggl(\frac{E_1}{c^2}  \biggl) \delta \phi \biggl] \biggl[ 1 - \alpha \biggl( \frac{\delta \phi}{c^2} \biggl) \exp(i \tilde \alpha) \biggl] \biggl[ 1 - \alpha \biggl( \frac{\delta \phi}{c^2} \biggl) \exp(- i \tilde \alpha) \biggl]
\nonumber\\
+\biggl[ E_2 + m_e \delta \phi + \biggl(\frac{E_2}{c^2}  \biggl)
\delta \phi \biggl] \biggl[ \exp(i \tilde \alpha) + \alpha \biggl(
\frac{\delta \phi}{c^2}\biggl) \biggl]  \biggl[ \exp(- i \tilde
\alpha) + \alpha \biggl( \frac{\delta \phi}{c^2}\biggl) \biggl\}.
\end{eqnarray}
As a result we obtain:
\begin{equation}
\biggl< E \biggl> = \frac{E_1 + E_2}{2} + \biggl( m_e + \frac{E_1
+ E_2}{2 c^2} + \frac{V_{12}}{c^2} \cos(\tilde \alpha) \biggl)
\delta \phi,
\end{equation}
where the expectation value of the gravitational mass per one
electron can be define as
\begin{equation}
\biggl< m_g \biggl> =  m_e + \biggl(\frac{E_1 + E_2}{2 c^2}\biggl)
+ \biggl(\frac{V_{12}}{c^2} \biggl) \cos(\tilde \alpha) ,
\end{equation}
Let us discuss Eq.(22). The first term is the bare electron mass,
the second term is the expected in relativity contribution to the
gravitational mass from the energies of electron levels (5). The
third term is the anomalous one and completely unexpected in
relativity contribution from the so-called virial term [6-8]. It
depends on the fixed difference of phases, $\tilde \alpha =
const$, in the coherent macroscopic ensemble of two stationary
states (5),(6). On the basis of Eq.(22), we can make conclusion
about the breakdown of the equivalence between expectation values
of energy and gravitational mass for the described above
Gravitational demon. We put forward a natural hypothesis that our
results are not restricted by ensembles of the hydrogen atoms but
are qualitatively true for any coherent macroscopic ensemble of
two or several stationary states of any composite quantum object.

In conclusion, we discuss in a brief some of the experimental
aspects of possible discovery of the violation of the Einstein's
Equivalence Principle. First of all, we again pay attention that
the phase difference in two quantum states in  the macroscopic
ensemble, $\tilde \alpha$, has to be fixed with good enough
accuracy (5),(6), otherwise the anomalous term in Eq.(22)
disappears after averaging over the phase. Such ensemble, in
principle, can be created by some laser technique with possibly
ultra-cold atoms or molecules in the Earth's laboratory [9]. The
second important point is that the measurements of a weight of the
ensemble has to be done very quickly, i.e. quicker than the time
scales characterizing wave functions in the quantum ensemble. Our
next steps will be to find more convenient composite quantum
objects for the above mentioned procedures.

We are thankful to Natalia N. Bagmet (Lebed), Steven Carlip,
Fulvio Melia, Pierre Meystre, Keneth Nordtvedt, Douglas Singleton,
and Vladimir E. Zakharov for useful discussions.

\end{document}